\documentclass[12pt]{iopart}
\usepackage{iopams}  
\usepackage{graphicx}% Include figure files
\begin{document}

\title[Saturable nonlinearity-induced mode-locked laser self-starting]{Effect of saturable nonlinearity on cw stability in passively mode-locked lasers with fast saturable absorbers}
\author{Alain M. Dikand\'e$^1$\footnote{corresponding author:\ead{dikande.alain@ubuea.cm}} and 
P. Achankeng Leke$^1$}
\address{$^1$ Laboratory of Research on Advanced Materials and Nonlinear Sciences (LaRAMaNS), Department of Physics, Faculty of Science, University of Buea P.O. Box 63, Buea Cameroon}
\vspace{10pt}
\begin{indented}
\item[]
\end{indented}

\begin{abstract}
The self-starting dynamics of a model for passively mode-locked lasers with saturable absorber, in which the optical amplifier has a saturable nonlinearity, is examined. The basic assumption is that the laser will operate in the mode-locked regime when the continuous-wave regime becomes unstable. Within the framework of the modulational-instability analysis, a global map for the laser self-starting criteria is constructed. According to this map, the saturable nonlinearity enhances the input-field intensity required for laser self-starting. Analytical expression of the zero modulation-frequency regime of this threshold input field is derived, which turns out to be valid in the normal as well as in the anomalous dispersion regime, where evidence of self-starting is also given.
\end{abstract}

%\pacs{}
\noindent{\it Keywords}: Operation regimes of laser optical systems, modulation and mode locking, optical systems with saturable nonlinearities
%\submitto{\LPL}
%\maketitle

\section{Introduction}
\large Mode-locked fiber lasers with saturable absorbers have become a particularly attractive source of high-intensity short pulses \cite{haus1,haus2,haus3,gordon1,gordon2,akhmed1,menyuk1,keller,kalash1,tang1,dik1,dik2}, for use in a vast area of modern communication technology. In these optical devices, an intensity fluctuation acts in conjunction with the fiber nonlinearity to modulate the cavity loss without need of some external control. Although several mode-locking techniques have been reported, for applications involving ultrashort pulses produced at finite repetition rates passive mode-locking remains the most preferred.\\ Passive mode-locking rests essentially on an appropriate choice of the gain medium, in this regard rare-earth-doped fiber amplifiers have demonstrated high efficiency in femtosecond pulse multiplexing and soliton-train laser applications \cite{pan,efra,miyo,wang}. Indeed this specific class of fiber-based gains is characterized by long upper-state lifetimes, so long that the gain changes only slowly within the cavity roundtrip. In general, because of the slow gain change, a fast saturable absorber will be required to clean up both the leading and the trailing edges of the pulse. Erbium-doped fibers \cite{pan,efra,miyo,wang} are the most appreciated choice among rare-earth-doped fiber amplifiers, they consist of silica optical fibers doped with rare earth ions (Er$^{3+}$), where the core of the amplifier is smaller than in typical fibers so as to increase the available density of erbium ions thus decreasing the optical pump threshold. These optical fibers can be transformed into laser amplifiers by simply adding positive feedback mechanisms, in this way erbium-doped fiber amplify optical signals by means of stimulated emission inducing a population inversion, where erbium ions are raised from their natural ground state to a higher energy level \cite{agar1}. Doping also provides means for controlling nonlinearity of the amplifier, which can be tuned from weak to strong \cite{iron1}. \\
Though passively mode-locked fiber lasers can display a wealth of operation regimes including continuous-wave (cw), plane-wave, period-doubling cascade, soliton and chaotic regimes \cite{akhmed1,kalash1,tang1,dik1,dik2,agar1,zhao1,kuntz1,tang2,yang2,zhao3,villa1}, in most applications it is desired that the device operates in the pulse regime. In this respect, a fundamental issue in recent studies of passively mode-locked lasers with saturable absorber has been their operation regime. As pioneer on this issue Haus proposed \cite{haus1} two possible schemes, the first involves direct simulations of the entire evolution of the optical light starting from noise, while the second focuses on the evolution of light intensity from cw. The second scheme rests on a master equation describing the propagation of the optical field in the laser cavity over roundtrips, thus allowing one follow its evolution from cw to pulse. If numerical simulations provide details about the evolution of the optical field from noise to either stable pulse or chaotic structures \cite{kiv1,kiv2}, it does not permit a global view of the field evolution over a relatively broad range of values of characteristic parameters of the laser system. On the contrary the second scenario, which is more analytical, puts into play the modulation of an input field of arbitrary intensity over the cavity roundtrips. Most importantly it has the merit to involve the cw as one of the transient regimes preceeding the fully pulsed regime, a feature reminiscent of the phenomenon of modulational instability suggesting that the cw will grow over roundtrips in the laser cavity, and should become unstable above some threshold intensity.\\
The self-starting mechanism of passively mode-locked lasers has been discussed previously \cite{menyuk1} by considering a complex Ginzburg-Landau equation (CGLE) with cubic nonlinearity, coupled to a two-level type dynamic gain. However, for applications involving pulse multiplexes, higher-order nonlinear terms and saturable nonlinearities in general \cite{hickman,hick1} are required to favor multi-periodic solitons and bound-soliton states\cite{kalash1,zhao1,kuntz1,tang2,zhao3,zhao2}. Therefore in this study we shall examine the self-starting dynamics of a mode-locked laser with saturable absorber as well as a saturable nonlinearity of the active medium. We follow the modulational instability analysis which, as stressed above, provides a global map of the dynamics of the laser system in different (i.e. cw and pulse) regimes of operation. In this approach, the laser is assumed to self-start when the cw regime is unstable and the laser operates in the mode-locked regime. Therefore, the main objective of this work is to investigate the effect of saturable nonlinearity on the self-starting feature of the family of mode-locked lasers considered. 
\section{Steady-state cw solution}
\label{sectwo}
\begin{figure}\centering
\includegraphics{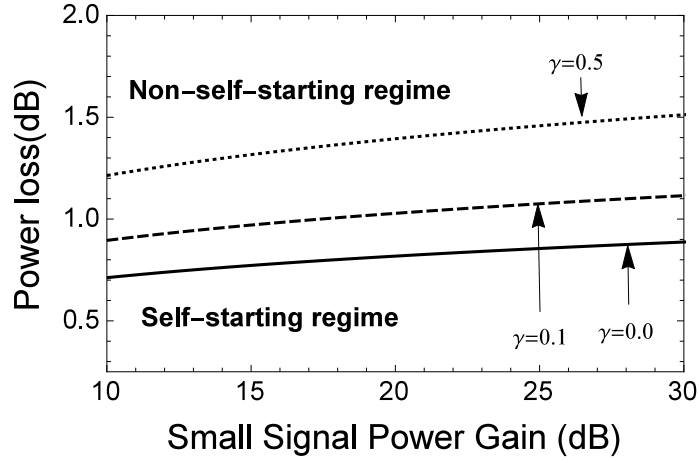}
\caption{Self-starting region in the $g_0$-$\ell$ plane for $P_c=1$, $P_s=2$, $\Gamma=0.01$ and three different values of the nonlinearity saturation coefficient namely $\gamma=0$, 0.1 and 0.5.}
\label{fig:one}
\end{figure}

We are interested in the self-starting dynamics of a passively mode-locked laser with fast saturable absorber, for which the optical amplifier has a saturable nonlinearity. Typical examples of optical amplifiers with saturable nonlinearity are semiconductor-doped glass fibers, indeed doping in these materials leads to non-cubic and sometimes extremely high optical nonlinearities \cite{hickman} as compared with conventional optical amplifiers with Kerr (i.e. cubic) nonlinearity \cite{menyuk1}. \\
The propagation of the laser field is governed by the following saturable-nonlinearity CGLE: 
\begin{equation}
\frac{\partial U}{\partial z}= \left(g - \ell + i\theta\right)U + \left(B + iD\right)\frac{\partial^2 U}{\partial t^2} + \frac{\Gamma +iK}{1+\gamma \vert U\vert^2}\vert U\vert^2 U, \label{mast}
\end{equation}
where $U(z,t)$ is the optical field, $z$ is the cavity roundtrip number, $t$ is time, $g$ is the gain, $\ell$ is the constant loss and $\theta$ is the phase change over each roundtrip. The characteristic parameters $C$ and $D$ account respectively for the spectral filter and group-delay dispersion, $\Gamma$ and $K$ are the fast saturable absorber and nonlinearity coefficients respectively, and $\gamma$ accounts for nonlinearity saturation in the fiber amplifier. The gain dynamics will be described by the following equation \cite{kart,haus4}:
\begin{equation} 
\frac{dg}{dt}=-\frac{(g-g_0)}{T_0}-\frac{|U(z,t)|^2}{T_0 P_s}g, \label{g0}
\end{equation}
where $g_0$ is the homogeneous gain, $P_s$ is the saturation power of the saturable absorber and $T_0$ is the gain relaxation time. \\
When $\gamma=0$, the above model reduces to the case discussed by Chen et al. in ref. \cite{menyuk1}. For small $\gamma$, the last term in eq. (\ref{mast}) can be expanded. This leads among others to the CGLE with a cubic-quintic nonlinearity \cite{akhmed4,akhmed2,akhmed3}. The self-starting dynamics in the case when the fiber amplifier has a cubic nonlinearity has been investigated by Chen et al. \cite{menyuk1}, within the framework of the modulational-instability approach. In the present study we shall examine the effect of nonlinearity saturation on self-starting, starting with the analysis of characteristic features of plane-wave solutions in the steady state. In this context solutions to the coupled set (\ref{mast})-(\ref{g0}) are given by:
\begin{equation}
U(z)= \sqrt{P_c}\,e^{i q_s z}, \hskip 0.5truecm g(t)= g_s, \label{cw}
\end{equation}
where $P_c=U_c^2$, $q_s$ is the plane-wave wave-number and $g_s$ is the steady-state gain. Replacing these in eqs. (\ref{mast}) and (\ref{g0}) and separating real from imaginary parts, we obtain:
\begin{equation}
g_s= \ell - \frac{\Gamma P_c}{1+\gamma P_c} = \frac{g_0}{1 + P_c/P_s}, \label{gsa}
\end{equation}
\begin{equation}
q_s= \theta + \frac{K P_c}{1+\gamma P_c}. \label{qs}
\end{equation}
Eqs. (\ref{gsa}) and (\ref{qs}) determine the steady-state gain $g_s$ and the wave-number $q_s$ for which the laser has a plane-wave form. Eq. (\ref{gsa}) is particularly relevant since according to eq. (\ref{mast}), the existence of plane wave will depend on the balance between the gain $g$ and the loss $\ell$. It follows from eq. (\ref{gsa}) that this balance should be determined by $P_c$ and $\gamma$, for a given saturation power $P_s$ of the saturable absorber. Fig. \ref{fig:one} summarizes the laser self-starting in the steady state regime, where the small signal power gain is defined as in ref. \cite{menyuk1} by $exp(2g_0)$ and where we considered three different values of $\gamma$, namely $\gamma=0$, $\gamma=0.1$ and $\gamma=0.5$. 

\section{Continuous-wave stability}
\label{secthree}
\begin{figure*}\centering
\begin{minipage}[c]{0.5\textwidth}
\includegraphics[width=2.9in,height=2.6in]{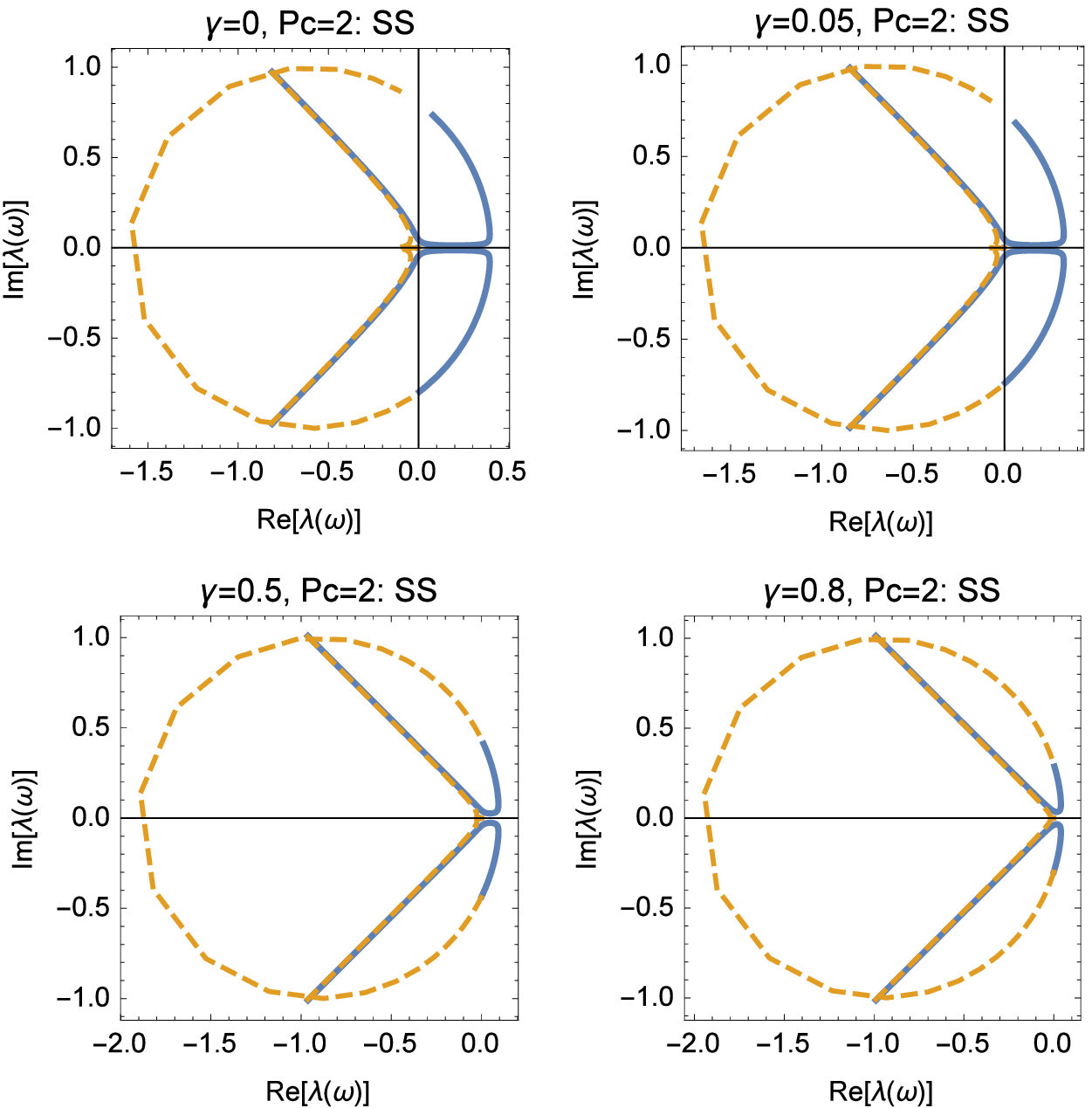}
\end{minipage}%
\begin{minipage}[c]{0.5\textwidth}
\includegraphics[width=2.9in,height=2.6in]{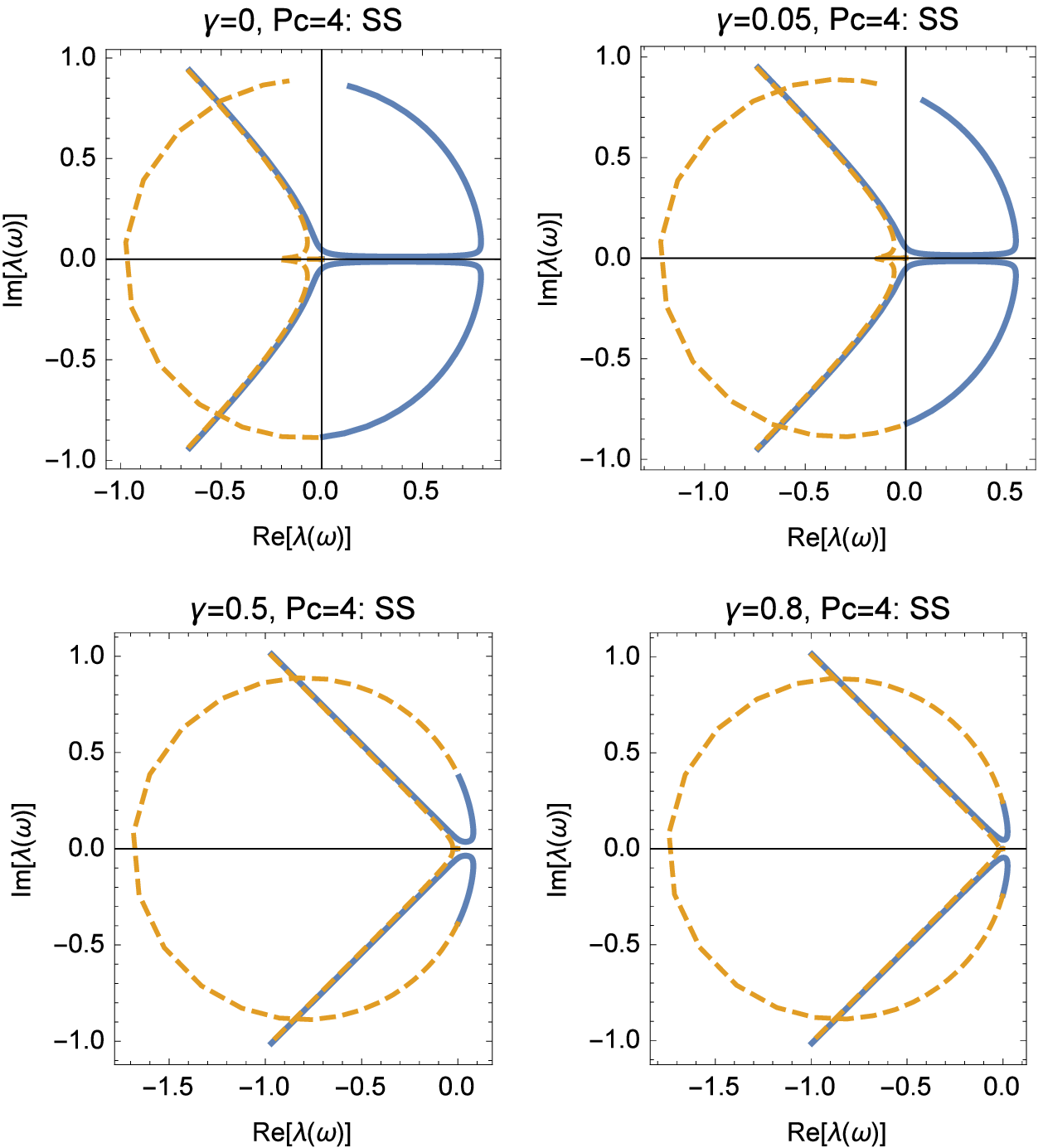}
\end{minipage}
\caption{(Color online) Parametric plots of $\kappa_1$ (solid curves) and $\kappa_2$ (dashed curves) in the complex plane, for $B=D=0.05$, $\Gamma=0.1$, $K=0.01$, $Ps=2$, $g_0= 4$, $T=200$ and different values of $\gamma$. Left two columns: $P_c=2$, right two columns: $P_c=4$. SS means "self-starting".}
\label{fig:three}
\end{figure*}

To investigate the stability of the cw field $U(z)$ given in formula (\ref{cw}), we carry out a modulational-instability analysis by following the evolution of the cw as it co-propagates with a plane-wave noise. Thus, consider a small perturbation $\tilde{u}(z,t)$ to the cw and a small deviation $\tilde{g}$ from the steady-state gain $g_s$, such that solutions to eqs. (\ref{mast}) and (\ref{g0}) now read: 
\begin{eqnarray}
 U(t)&=& \left[\sqrt{P_c} + \tilde{u}(z,t)\right]e^{(iq z)}, \nonumber  \\
 g(t)&=& g_s + \tilde{g}. \label{ss1}
\end{eqnarray}
Replacing these in eqs. (\ref{mast}) and (\ref{g0}) and linearizing, we find:
\begin{eqnarray}
\tilde{u}_z &=& \alpha \tilde{u}_{tt} + \left(\tilde{u} + \tilde{u}^*\right)F_u + \sqrt{P_c}\tilde{g}, \label{ueqt} \\
\tilde{g}_t&=& -\frac{\tilde{g}}{T_e} + F_g(\tilde{u}, \tilde{u}^*),  \label{geqt}
\end{eqnarray}
with 
\begin{eqnarray}
\alpha&=& B+iD, \hskip 0.5truecm F_u=P_c\frac{\Gamma + iK}{(1+\gamma P_c)^2}, \nonumber \\
F_g&=& -\frac{\epsilon_c}{T_e}(\tilde{u}+\tilde{u}^*), \hskip 0.2truecm \epsilon_c= \frac{g_0\sqrt{P_c}}{P_s(1+P_c/P_s)^2},  \label{geqa}       
\end{eqnarray}
and the effective relaxation time $T_e$ is defined as:
\begin{equation}
T_e= \frac{T_0}{1+P_c/P_s}.
\end{equation}
The linear inhomogeneous first-order ordinary differential equation eq.~(\ref{geqt}) can be solved by means of the Green-function method yielding:
\begin{equation}
 \tilde{g}(t)=\int_{-\infty}^t{G(t,t')F_g[\tilde{u}(t'), \tilde{u}^*(t')]dt'},
\end{equation}
where $G(t,t')= e^{-(t-t')/T_e}H(t-t')$ is the Green function with $H(t-t')$ the step function. Now assuming:
\begin{equation}
[\tilde{u}(z,t), \tilde{u}^*(z,t)]= [A_1, A_2] e^{(\kappa z + i\omega t)},  \label{cw1}
\end{equation}
where $\kappa$ is the rate of spatial amplification of the noise and $\omega$ the associate time modulation frequency, eqs.~(\ref{ueqt}) together with its complex conjugate lead to the following secular equation in matrix form:
\begin{equation}\label{egen1}
\kappa \left(
\begin{array}{c}
A_1 \\A_2
\end{array}\right)= \left [\left(
                    \begin{array}{cc}
                        m_1 & m_2\\
                        m_2^{*} & m_1^{*}

                        \end{array}
                        \right)-m_0
                        \left(
                        \begin{array}{cc}
                            1&1 \\
                            1&1
                            \end{array}\right)\right]\; \left(
                        \begin{array}{cc}
                       A_1 \\ A_2
                        \end{array}
                        \right),
                        \end{equation}    
with:
\begin{equation}
 m_1=-\alpha + F_u, \hskip 0.5truecm m_2= F_u, \hskip 0.5truecm m_0= \frac{\sqrt{P_c}\epsilon_c}{1+i\omega T_e}. \label{solut2}
\end{equation}
The determinant of the above $2\times 2$ matrix gives rise to a quadratic polynomial in the associate eigenvalue $\kappa$, the two possible roots of which read:
\begin{eqnarray}
 \kappa_{1,2} &=& \frac{\Gamma P_c}{(1+\gamma P_c)^2} - B\omega^2 - m_0 \nonumber \\
 &\pm& \sqrt{\left[m_0 - \frac{\Gamma P_c}{(1+\gamma P_c)^2}\right]^2 - (D\omega^2)^2 + \frac{2D K P_c\omega^2}{(1+\gamma P_c)^2}}, \nonumber \\
 \label{solut}
\end{eqnarray}
where the subscripts 1, 2 refer to the plus and minus signs, respectively. According to formula (\ref{solut}), the cw will be unstable (and hence the laser will self-start) if the real part of $\kappa$ is positive. It turns out that at zero modulation frequency, when $\kappa_1= 0$ and $\kappa_2=2\frac{\Gamma P_c}{(1+\gamma P_c)^2} - 2\sqrt{P_c}\epsilon_c$, we need $\epsilon_c < \frac{\Gamma \sqrt{P_c}}{(1+\gamma P_c)^2}$ for the laser to self-start. Quantitatively, this condition implies two characteristic values of $P_c$ above which self-starting can occur. One of them is negative and hence unstable, while the positive one,
\begin{equation}
 P_c^{(\gamma)}=\frac{P_c^{(0)}}{1 - \gamma \sqrt{P_s g_0/\Gamma}}, \label{solut1}
\end{equation}
sets a threshold input-field intensity above which the laser will self-start. Note that in the case of mode-locked lasers with a cubic-nonlinearity optical amplifier this threshold value should be:
\begin{equation}
P_c^{(0)}=P_s\left(\sqrt{g_0/P_s\Gamma} - 1\right). \label{cub}
\end{equation}
For a best understanding of laser self-starting when the two eigenvalues vary with the modulation frequency $\omega$, we resorted to a global mapping of the $\kappa_{1,2}$ in the complex plane. Thus, fig. \ref{fig:three} are parametric plots of the two eigenvalues for the modulation frequency in the range $-5\leq\omega\leq 5$, where the imaginary part $Im(\kappa)$ is plotted as a function of the real part $Re(\kappa)$. Values of characteristic parameters of the model are indicated in the caption.

\begin{figure*}\centering
\begin{minipage}[c]{0.5\textwidth}
\includegraphics[width=2.9in,height=2.6in]{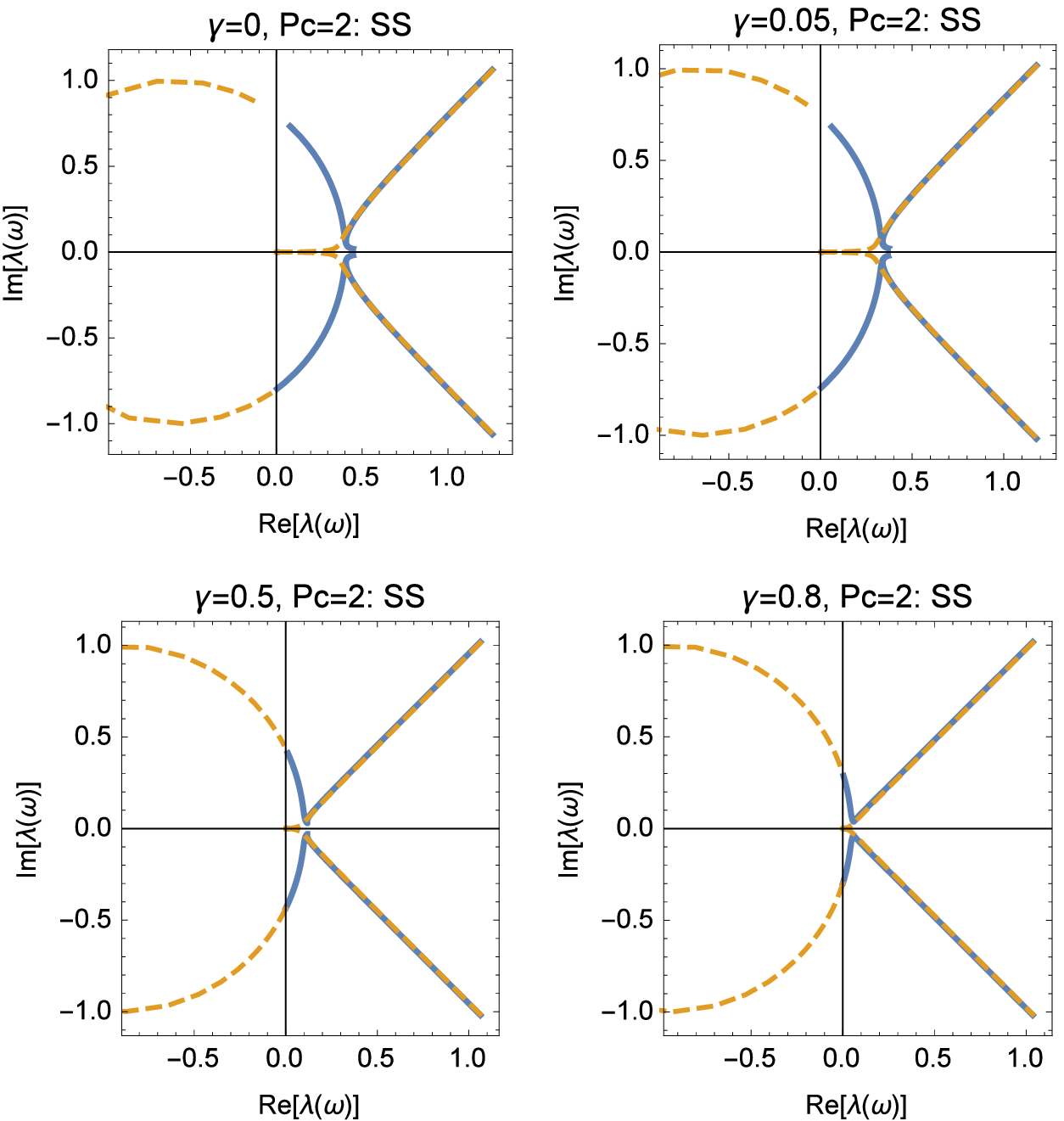}
\end{minipage}%
\begin{minipage}[c]{0.5\textwidth}
\includegraphics[width=2.9in,height=2.6in]{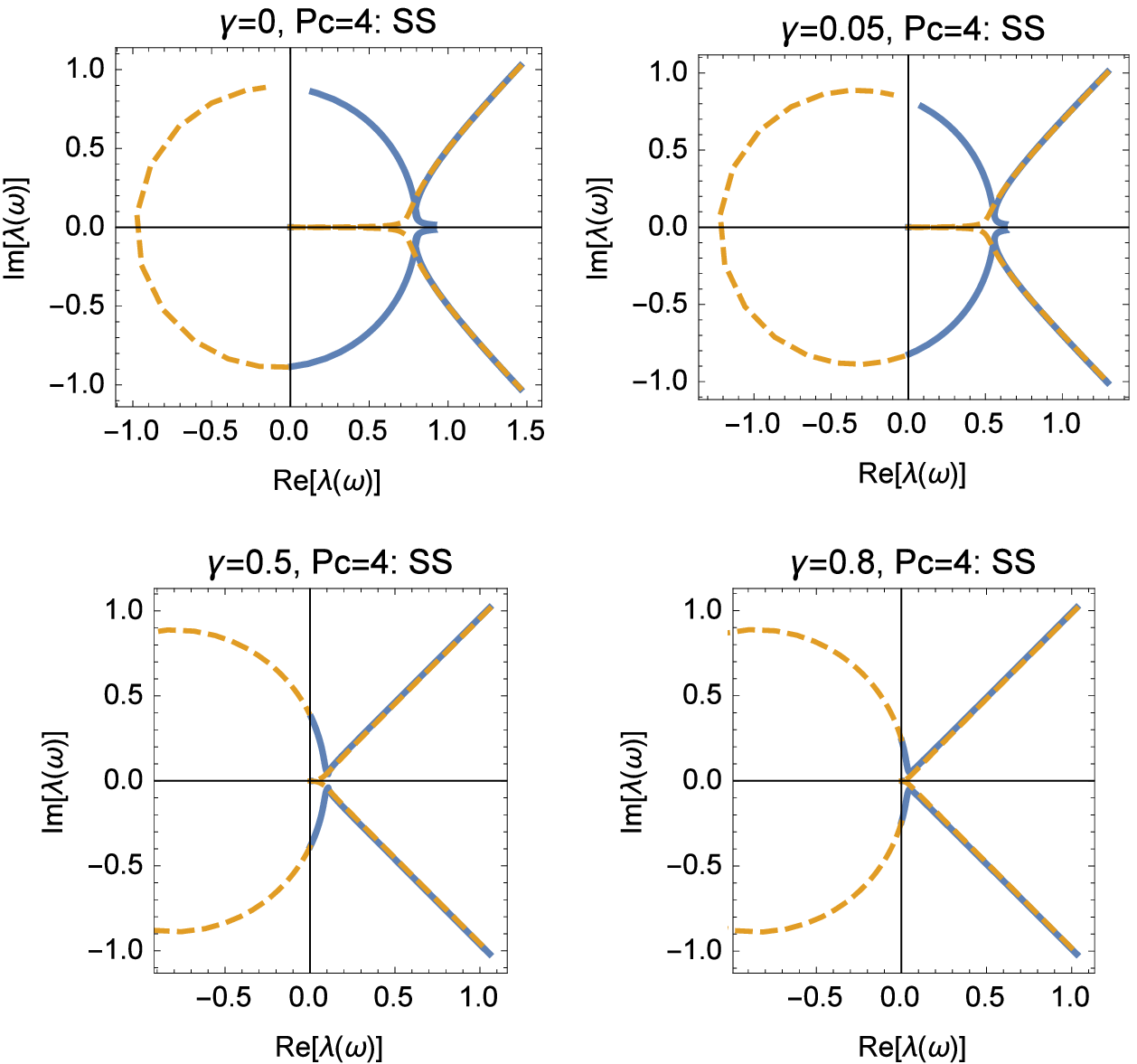}
\end{minipage}
\caption{(Color online) Parametric plots of $\kappa_1$ (solid curves) and $\kappa_2$ (dashed curves) in the complex plane, for $B=D=-0.05$, $\Gamma=0.1$, $K=0.01$, $Ps=2$, $g_0= 4$, $T=200$ and different values of $\gamma$. Left two columns: $P_c=2$, right two columns: $P_c=4$.}
\label{fig:four}
\end{figure*}

 According to the graphs, self-starting is stronger for the cubic nonlinearity and is enhanced by an increase of $P_c$. However, when we increase $\gamma$ for a fixed value of the input intensity $P_c$, the cw field bedomes highly unstable. Actually this later observation is consistent with the dependence of the threshold value $P_c^{(\gamma)}$ of $P_c$, on the nonlinearity saturation coefficient $\gamma$ obtained in formula (\ref{solut1}) and suggesting a higher input field for laser self-starting when the nonlinearity is of a saturable type. \\
It is worthwile recalling that the above analysis assumes the laser will self-start (automatically in the pulse regime) when the cw regime is unstable. On the other hand in our discussions we assumed that the optical system is in a normal dispersion regime (i.e. B and D are positive), where the CGLE is equivalent to the NLSE such that the system admits quasi-Schr\"odinger sech-type pulses in the mode-locked regime. Still, although negative group-delay dispersion and spectral filter act against NLS sech-type pulses, experiments have demonstrated that pulses can still form in this case. The most interesting experimental evidences have been reported in ref. \cite{kalash1}, where multi-periodic and bound pulse states have been shown to form when $B$ and $D$ cross zero from the positive branch. As the two parameters decrease in the negative branch, multiple-pulse structures sharpen while pulse durations get shorter and shorter. So to say self-starting is also possible in the anomalous dispersion regime, and in the present particular context this is evidenced by the parametric plots of $\kappa_1$ and $\kappa_2$ shown in fig. \ref{fig:four}. 

\section{Conclusion}
\label{secfour}
The present work is to be seen as an extension of the study carried out in ref. \cite{menyuk1}, to the general context of mode-locked lasers with highly nonlinear optical amplifiers. In this respect, while we assume, as in this previous work, that the laser will self-start (i.e. operate instantaneously) once the cw regime is unstable, the modulational-instability analysis can actually only provide relevant information about the stability of the cw regime. It strictly cannot help identiy the exact structures of the pulse and/or multiple-pulse fields that would be stabilized in the mode-locked regime. To determine the exact structures of the laser field in the mode-locked regime we must solve the CGLE. In this task, numerical simulations will enable us explore all the possible operation regimes inherent to the dynamics of the above model. This is expected to reveal not just stable pulse or multipulse regimes, but also transient states characterized by period-doubling cascades and chaotic phases as established for the cubic-quintic CGLEs \cite{akhmed4,akhmed2,akhmed3,akhmed5}. 

\ack
The work of A. M. Dikand\'e is supported by the Alexander von Humboldt foundation.


\section*{References}
\begin{thebibliography}{99}
\bibitem{haus1} Haus H. A. 1975 J. Appl. Phys. \textbf{46}, 3049. 
\bibitem{haus2} Haus H. A. and Silberberg Y. 1986 IEEE J. Quantum Electron. \textbf{22}, 325.
\bibitem{haus3} Ippen  E. P., Haus H. A. and Liu L. Y. 1989 J. Opt. Soc. Am. \textbf{B6}, 1736.
\bibitem{gordon1} Martinez O. E., Fork R. L. and Gordon J. P. 1984 Opt. Lett. \textbf{9}, 156.
\bibitem{gordon2} Well R., Vodonos R. B., Gordon A., Gat O. and Fischer B. 2007 Phys. Rev. \textbf{E76}, 031112.
\bibitem{akhmed1} Akhmediev N. N., Lederer M. J. and Luther-Davies B. 1998 Phys. Rev. \textbf{E57}, 3664.
\bibitem{menyuk1} Chen C. J., Wai P. K. A. and Menyuk C. R. 1994 Opt. Lett. \textbf{20}, 350.
\bibitem{keller} Pschotta R. and keller U. 2001 Appl. Phys. \textbf{B73}, 653.
\bibitem{kalash1} Kalashnikov V. L. , Sorokin E. and Sorokina I. T. 2003 IEEE J. Quant. Elec. \textbf{39}, 323.
\bibitem{tang1} Tang D. Y., Zhang H., Zhao L. M. and Wu X. 2008 Phys. Rev. Lett. \textbf{101}, 153904.
\bibitem{dik1} Fandio Jubgang Jr. D., Dikand\'e A. NM. and Sunda-Meya A. 2015 Phys. Rev. \textbf{A92}, 053850.
\bibitem{dik2} Fandio Jubgang Jr. D. and Dikand\'e A. M. 2017 J. Opt. Soc. Am. \textbf{B34}, 2721.
\bibitem{pan} Pan C. L., Hwang C. D., Kuo J. C., Shieh J. M. and Wu K. H. 1992 Opt. Lett. \textbf{17}, 1444.
\bibitem{efra} Beltr\'an E. M., Selected Topics on Optical Fiber Technology, (Ed. Yasin M., InTech Publishing, China, 2012).
\bibitem{miyo} Miyoshi T., Makidera M., Kawamura T., Kashima S., Matsuo S. and Kaneda T 2002 Jpn. J. Appl. Phys. \textbf{41}, 5262.  
\bibitem{wang} Wang Q., Geng J., Luo T. and Jiang S. 2009 Opt. Lett. \textbf{34}, 3616.
\bibitem{agar1}Agarwal G. P., Nonlinear Fiber Optics (1$^{rst}$ed., Academic, Boston, Mass., 1989).
\bibitem{iron1} Ironside C. N., Cullen T. J., Bhumbra B. S., Bell J., Banyai W. C., Finlayson N., Seaton C. T. and Stegeman G. I. 1988 J. Opt. Soc. Am. \textbf{B5}, 492.
\bibitem{zhao1} Zhao B., Tang D. Y., Shum P., Man W. S., Tam H. J., Gong Y. D. and Lu C. 2004 Opt. Commun. \textbf{229}, 363.
\bibitem{kuntz1}Li. F., Wai P. K. A. and Kutz J. N. 2010 J. Opt. Soc. Am. \textbf{B27}, 2068.
\bibitem{tang2} Tang D. Y., Zhao L. M. and Li F. 2005 Europhys. Lett. \textbf{71}, 56.
\bibitem{yang2} Yang L., Zhang L., Yang R., Yang L., Yue B. and Yang P. 2012 Opt. Commun. \textbf{285}, 143.
\bibitem{zhao3} Zhao L. M., Tang D. Y. and Liu A. Q. 2006 Chaos \textbf{16}, 013128.
\bibitem{villa1} Villanueva G. E. and P\'erez-Mill\'an P. P. 2012 Opt. Lett. \textbf{37}, 1971.
\bibitem{kiv1}If F., Christiansen P. L., Elgin J. L., Gibbon J. D. and Skovgaard O. 1986 Opt. Commun. \textbf{57}, 350.
\bibitem{kiv2} K. J. Blow and D. Wood, J. Opt. Soc. Am. \textbf{B5} (1988) 629.
\bibitem{hickman}Hickmann J. M., Cavalcanti S. B., Borges N. M., Gouveia E. A. and Gouveia-Neto A. S. 1993 Opt. Lett. \textbf{18}, 182.
\bibitem{hick1}Lyra M. L. and Gouveia-Neto A. S. 1994 Opt. Commun. \textbf{108}, 117.
\bibitem{zhao2}Zhao M. L., Tang D. L. and Zhao B. 2005 Opt. Commun. \textbf{252}, 167.
\bibitem{kart} K\"artner F. X., Jung I. D. and Keller U 1996 IEEE J. Selec. Topics Quantum Electron. \textbf{2}, 540.
\bibitem{haus4} Haus H. A., Ippen E. P. and Tamura K. 1994 IEEE J. Quantum Electron. \textbf{30}, 200.
\bibitem{akhmed4}Soto-Crespo J. M. and Akhmediev N. N. 1999 J. Opt. Soc. Am. \textbf{B16}, 674.
\bibitem{akhmed2}Soto-Crespo J. M., Akhmediev N. N.  and Town G. 2002 J. Opt. Soc. Am. \textbf{B19}, 234.
\bibitem{akhmed3}Soto-Crespo J. M., Grapinet M., Grelu P. and Akhmediev N. N. 2004 Phys. Rev. \textbf{E70}, 066612.
\bibitem{akhmed5} Akhmediev N. N., Afanasjev V. V. and Soto-Crespo J. M. 1996 Phys. Rev. \textbf{E53}, 1190.

\end{thebibliography}
\end{document}